\documentclass[onecolumn]{aastex63}
\usepackage{amsmath}
\usepackage{multirow}
\usepackage{hyperref}
\newcommand{\PMO}{Key Laboratory of Dark Matter and Space Astronomy, Purple Mountain Observatory, Chinese Academy of Sciences, Nanjing 210033, People's Republic of China}
\newcommand{\USTC}{School of Astronomy and Space Science, University of Science and Technology of China, Hefei, Anhui 230026, People's Republic of China}
\newcommand{\ITGU}{Institut f\"{u}r Theoretische Physik, Goethe Universit\"{a}t, Max-von-Laue-Str. 1, D-60438 Frankfurt am Main, Germany.}
\newcommand{\NNU}{Department of Physics and Institute of Theoretical Physics, Nanjing Normal University, Nanjing 210046, People's Republic of China}

\begin{document}
\title{Bulk properties of PSR J0030+0451 inferred with the compactness measurement of NICER}
\author[0000-0003-3471-4442]{Chuan-Ning Luo}
\affil{\PMO}
\affil{\USTC}
\author[0000-0001-9120-7733]{Shao-Peng Tang}
\affil{\PMO}
\email{Corresponding author.~tangsp@pmo.ac.cn}
\author[0000-0001-9034-0866]{Ming-Zhe Han}
\affil{\PMO}
\author[0000-0002-9078-7825]{Jin-Liang Jiang}
\affil{\ITGU}
\author[0000-0001-7821-3864]{Wei-Hong Gao}
\affil{\NNU}
\author[0000-0002-9758-5476]{Da-Ming Wei}
\email{Corresponding author.~dmwei@pmo.ac.cn}
\affil{\PMO}
\affil{\USTC}

\begin{abstract}
In 2019, Neutron star Interior Composition ExploreR (NICER) mission released its findings on the mass and radius of the isolated neutron star (INS) PSR J0030+0451, revealing a mass of approximately 1.4 solar masses ($M_{\odot}$) and a radius near 13 kilometers. However, the recent re-analysis by the NICER collaboration \citep{vinciguerra2024updated} suggests that the available data primarily yields a precise inference of the compactness for this source while the resulting mass and radius are strongly model-dependent and diverse (the 68.3\% credible regions just overlap slightly for the ST+PDT and PDT-U models). By integrating this  compactness data with the equation of state (EoS) refined by our latest investigations, we have deduced the mass and radius for PSR J0030+0451, delivering estimates of $M=1.48^{+0.09}_{-0.10}~M_\odot$ and $R=12.38_{-0.70}^{+0.51}~{\rm km}$ for the compactness found in ST+PDT model, alongside $M=1.47^{+0.14}_{-0.20}~M_\odot$ and $R=12.37_{-0.69}^{+0.50}~{\rm km}$ for the compactness in PDT-U model. These two groups of results are well consistent with each other and the direct X-ray data inference within the ST+PDT model seems to be favored. Additionally, we have calculated the tidal deformability, moment of inertia, and gravitational binding energy for this NS. Furthermore, employing these refined EoS models, we have updated mass-radius estimates for three INSs with established gravitational redshifts.
\end{abstract}

\section{Introduction}
Since the discovery of neutron stars \citep[NSs;][]{hewish1968observation}, astronomers have identified over 3,000 such entities. However, less than 5\% of those has a mass which has been accurately measured \citep[see][for a recent summary]{shao2020maximum}, with almost all these stars being part of binary systems. This is because the masses of NSs in binary systems can be estimated through some feasible approaches \citep[see a comprehensive review in][and references therein]{2016ARA&A..54..401O}. In contrast, measuring the masses of isolated neutron stars (INSs), which constitute the majority, poses substantial difficulties. The Neutron star Interior Composition ExploreR (NICER) team has advanced this field by utilizing X-ray pulse profile modeling technique to determine the mass of INSs, thereby providing a novel approach to the measurement of NS mass-radius relationships \citep{2019ApJ...887L..26B}. In 2019, the NICER collaboration published their findings on the INS PSR J0030+0451, revealing an estimated mass of approximately 1.4$M_{\odot}$ and a radius near 13 kilometers \citep{riley2019nicer, miller2019psr}.

However, recent studies by the NICER team suggest that these initial measurements require a reassessment \citep{vinciguerra2024updated}. Adopting more general assumptions and integrating data from the XMM-Newton detector, these authors have found that current data can only precisely constrain the ratio of mass to radius (the compactness of NS, ${\cal C}=M/R$) for PSR J0030+0451 rather than accurately determining its mass and radius separately. Within the X-PSI (X-ray Pulse Simulation and Inference) framework, four models with increasing complexity, namely, ST-U, ST+PST, ST+PDT, and PDT-U, were developed by these authors. 
The fitting results indicate that the latter two models yield more reliable $(M,~R)$ measurements. For ST+PDT and PDT-U models, they obtain $(M,~R)=(1.40^{+0.13}_{-0.12}~M_\odot,~11.71^{+0.88}_{-0.83}~{\rm km})$ and $(1.70^{+0.18}_{-0.19}~M_\odot,~14.44^{+0.88}_{-1.05}~{\rm km})$, respectively \citep{vinciguerra2024updated}. Note that in this work the error bars are for $68.3\%$ credible level. Clearly, these two sets of results are strongly model-dependent. 
Nevertheless, the inferred compactness ${\cal C}$ for both models closely aligns, each with a marginal uncertainty (ST+PDT model results in ${\cal C}=0.1773_{-0.0074}^{+0.0056}$; PDT-U model results in ${\cal C}=0.179_{-0.022}^{+0.011}$), and the relative difference between their central values is only about $1\%$.

In \citet{TangSP2020} and \citet{2022ApJ...930....4L}, a new method to infer the mass and radius of INSs with measured gravitational redshifts ($z_{\rm g}$) has been developed. The idea is that the effective constraints on the mass-radius ($M-R$) relationship can be derived from nuclear theories/experiments and multimessenger observations of NSs. These constraints facilitate the establishment of a corresponding $z_{\rm g}-M$ relationship and hence the inference of mass and radius distributions given $z_{\rm g}$ measurements. Rather similarly, a ${\cal C}-M$ relationship can be established, which can then be directly applied to PSR J0030+0451 to infer its mass and radius. 
To achieve this goal, the latest constraints on the EoS are required. Unfortunately, in the last two years, neither the LIGO/Virgo collaboration nor the NICER mission has provided new, improved NS mass-radius measurements. An improvement we present here is the incorporation of NS mass function information following \citet{Fan2023}, which provides further constraints on the equation of state (EoS). Unlike \citet{TangSP2020} and \citet{2022ApJ...930....4L}, which only took parameterized EoS representation/reconstruction, our approach is based on the EoS (non-)parametric reconstruction results using feed-forward neural network \citep[i.e., FFNN; see][]{2021ApJ...919...11H,han2023plausible}, flexible piecewise linear sound speed \citep[i.e., PWLS; see][]{jiang2023bayesian}, and Gaussian processes \citep[i.e., GP; see][]{2020PhRvD.101l3007L,2023ApJ...950..107G} methods. In this work, we utilize these newly constrained EoSs (see Section \ref{sec:methods} for the detail)
to infer the bulk properties of the INS PSR J0030+0451.

Our work is organized as follows. In Section \ref{sec:methods} we introduce the methods. The results of the calculation are presented in Section \ref{sec:results}. Section \ref{sec:discussion} is the conclusion and discussion. Throughout this work, the Newtonian gravitational constant and the speed of light are all set to $1$, i.e., $G=c=1$.
 
\section{Methods}\label{sec:methods}

\subsection{The data and information used in the EoS reconstruction}\label{subsec:data}
It is widely known that the nuclear physics data play a dominant role in governing the EoS at densities below $\sim 1n_{\rm s}$, where $n_{\rm s}$ is the nuclear saturation number density. 
For the density below $0.3n_{\rm s}$, we match the constructed EoS to the NS crust EoS \citep{Douchin:2001sv}.
At the densities of $0.3n_{\rm s}\leq n\leq 1.1n_{\rm s}$, the next-to-next-to-next-to-leading order $\chi$EFT calculation results \citep{Drischler:2017wtt} serve as the constraint, where we exclude the EoSs exceeding the 3$\sigma$ range of $\chi$EFT for the FFNN and PWLS methods, while for GP method we take it as the training data for conditioning the GP (see \citet{Tang2024} for more information). At high densities reachable in the neutron stars, the constraints are mainly contributed by the multimessenger observation data, including the first double neutron star merger event GW170817 \citep{GW170817-a} and the the mass-radius measurements of PSR J0740+6620 by NICER \citep{2021ApJ...918L..27R, 2021ApJ...918L..28M}. 
Note that for a given EoS, one can have a $M-R$ curve of the NSs via solving the famous Tolman–Oppenheimer–Volkoff equation. In turn, the measured $M,~R$ of a group of NSs can be used to reconstruct the EoS.
Unlike \citet{TangSP2020} and \citet{2022ApJ...930....4L}, now the NICER data for PSR J0030+0451 are no longer included in the constraint on the EoS. While for PSR J0740+6620, the results are taken into account because they are more robust, possibly due to its X-ray faintness, tighter external constraints (in particular its accurately measured mass, see \citet{Fonseca:2021wxt}), and/or viewing geometry \citep{2023ApJ...956..138S}.
At even higher densities, we can employ the extrapolated constraints set by the pQCD calculation that is only valid at 
very high densities of $\geq 40n_{\rm s}$ \citep{2021PhRvL.127p2003G}. One widely-adopted approach is to extrapolate the $\chi$EFT and NS data bounded EoS to a density of $\sim 10n_{\rm s}$ and then reject the models violating the pQCD constraints \cite[e.g.,][]{2020NatPh..16..907A,2022ApJ...939L..34A,han2023plausible,jiang2023bayesian}. In this work we follow such an approach. To strengthen the constraints on the EoS, we also incorporate the $M_{\rm TOV}$ information inferred from the NS mass function (see Sec. III.A of \citet{Fan2023} for the details).

\subsection{EoS reconstruction methods}\label{subsec:EoS}

Similar to \citet{TangSP2020} and \citet{2022ApJ...930....4L}, we can obtain the $(\cal C,M)$ information of NSs based on the constraint results of the EoS. We adopt the recently-developed FFNN method \citep{2021ApJ...919...11H,han2023plausible}, the PWLS model \citep{2022ApJ...939L..34A,jiang2023bayesian} and the GP method \citep{essick2020nonparametric,2023ApJ...950..107G,Tang2024}. Below we introduce these three methods briefly.

In the FFNN method \citep{han2023plausible},  the neutron star EoS is represented by the 10-node single-layer FFNN expansion that is capable of fitting theoretical EoSs very well \citep{2021ApJ...919...11H}, which is given by 
\begin{equation}
c_{\rm s}^2(\rho) = S\left(\sum_i^{\rm N} w_{2i}\sigma(w_{1i}\ln\rho+b_{1i})+b_{2}\right),
\end{equation}
where $\rho$ is the rest-mass density, ${\rm N}=10$, $c_{\rm s}^2={\rm d}p/{\rm d}\varepsilon$ is the squared sound speed, $\sigma(\cdot)$ is the activation function, $w_{\rm 1i}$, $w_{\rm 2i}$, $b_{\rm 1i}$, and $b_2$ are weights/bias parameters of the FFNN. S$(\cdot)$ is the sigmoid function, i.e., $S(x)=1/(1+e^{-x})$, which guarantees the microscopic stability and causality condition.
Two activation functions, including $1/(1+e^{-x})$ (i.e., the {\it sigmoid}) and ${(e^x-e^{-x})}/{(e^x+e^{-x})}$ (i.e., the {\it hyperbolic tangent}), have been adopted and the resulting two sets of posteriors are combined to obtain the results. It should be noted that the inclusion of a hyperbolic tangent activation function allows for the potential of a phase transition (PT) phenomenon.
The weights and bias parameters are uniformly sampled in $(-5, 5)$. The EoS with $M_{\rm TOV}$ beyond $1.4-3~M_\odot$ is discarded during the inference. We use the Bayesian inference library {\sc BILBY} \citep{2019ApJS..241...27A} with the sampling algorithm {\sc PyMultiNest} \citep{2014A&A...564A.125B} to obtain the posterior samples of the EoSs.

As for the PWLS method, below densities $0.5n_{\rm s}$ the EoS is described by the Baym-Bethe-Pethick (BPS) EoS \citep{1971NuPhA.175..225B}. At higher densities it is then divided into 11 segments. The first segment is represented by a single polytrope, while the others are characterized by linear segments of the chemical potential $\mu-c_{\rm s}^{2}$ relations, with parameters $c_{{\rm s},i}^{2}$ at logarithmically separated fixed chemical potential positions \citep{jiang2023bayesian}. The likelihood construction follows the same approach as the FFNN method to ensure comparability of results.

For the GP approach, following \citet{2023ApJ...950..107G} we take $\phi$ as a function of the baryon number density $n$ to describe the NS EoS, i.e., 
\begin{equation}
    \phi = -\ln(1/c_{\rm s}^2 -1).
\end{equation}
The GP of $\phi(n)$ can be described as 
$\phi(n) \sim \mathcal{N}(-\ln(1/\bar{c_{\rm s}}^2 -1), K(n_i, n_j))$,
where the kernel function of the GP is given by $K(n_i, n_j)=\eta^2 e^{-(n_i-n_j)^2/2l^2}$. The distributions of the three hyperparameters, namely the variance $\eta$, the correlation length $l$, and the mean speed of sound squared $\bar{c}_s^2$ are $\eta \sim \mathcal{N}(1.25,0.2^2)$, $l \sim \mathcal{N}\left(0.5n_s, (0.25n_s)^2\right)$, and $\bar{c}_s^2 \sim \mathcal{N}(0.5,0.25^2)$, respectively.
The EoS posterior of GP method is selected according to their likelihoods/weights from the generated samples.
Additionally, The GP and PWLS method also possess the capability of generating EoS that exhibit PT characteristics in principle.

\subsection{The bulk properties of the neutron star as a function of the compactness}
Incorporating the data/information summarized in Section \ref{subsec:data} into the above reconstruction methods, the bounded EoSs can be inferred via the Bayesian inference. In Figure \ref{fig:MC_compare} we show the resulting ${\cal C}-M$ relationship (see also the $z_g-M$ relation in Appendix~\ref{app:zg-m}). The three regions marked in {red, blue and green represent the results of the GP method, FFNN method, and PWLS method, respectively.} Evidently, these three regions are almost identical, suggesting that the results are not sensitive to the EoS reconstruction methods. Below we will present the relationships between the compactness of a non-rotating NS and the tidal deformability $\Lambda$, the moment of inertia $I$ and the gravitational binding energy ${\rm BE}$, respectively. For simplicity, we just take the Gaussian process result to carry out the following calculations.  

\begin{figure}
    \centering
    \includegraphics[width=0.6\textwidth]{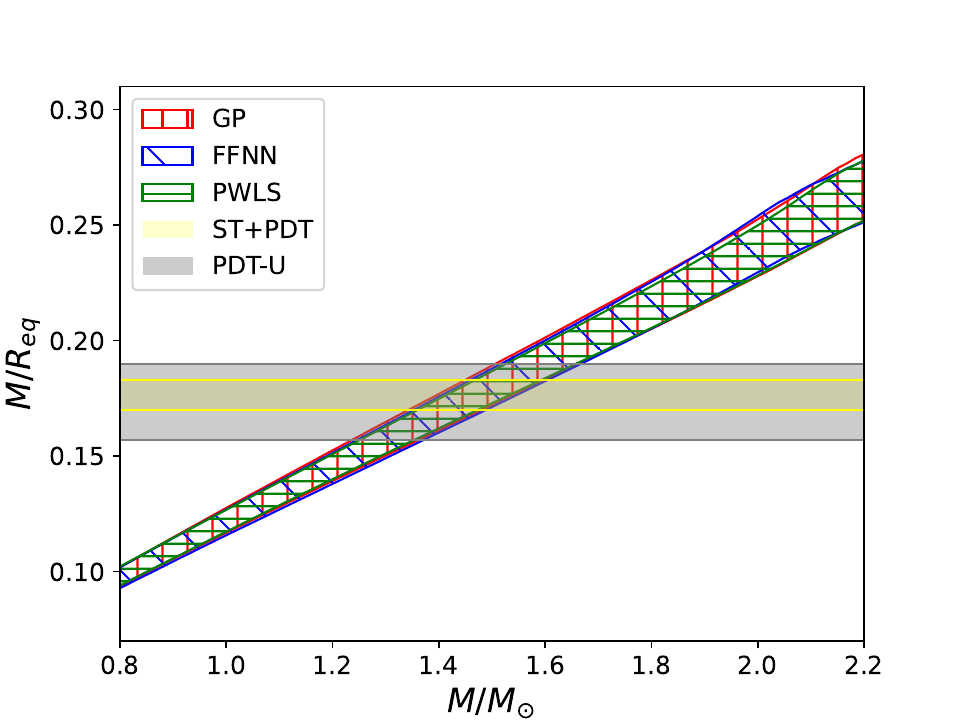}
    \caption{The ${\cal C}-M$ relationship diagram. The red area represents the results obtained using the GP model, the blue area represents the results obtained using the FFNN method, the green area represents the results obtained using the PWLS method, and the yellow and gray filled areas respectively represent the results from the ST+PDT and PDT-U models used by \citet{vinciguerra2024updated}. The regions are for the $68\%$ credible level.}
    \label{fig:MC_compare}
\end{figure}

The ability of a star being distorted by the gravitational field from its companion star is quantified by the tidal deformability.  
This tidal deformability $\lambda$ is defined as \citep{thorne1998tidal,2020GReGr..52..109C}
\begin{equation}
    \lambda = -\frac{Q_{ij}}{\varepsilon_{ij}},
\end{equation}
where $Q_{ij}$ is the quadrupole moment of the star and 
$\varepsilon_{ij}$ is the gravitational field.
The dimensionless form is
\begin{equation}
    \Lambda = \frac{\lambda}{M^5} = \frac{2}{3} k_2 \frac{R^5}{M^5} = \frac{2}{3} k_2 C^{-5},
\end{equation}
where $k_2$ represents the Love number, and $M$ and $R$ are the NS's gravitational mass and radius, respectively. To compute $\Lambda$, one can solve the Regge-Wheeler equation, yielding $\Lambda$ as a function of the compactness $C = M/R$ and a surface value $Y$ that measures the relativistic quadrupole gravitational potential induced by the tidal deformation \citep{hinderer2008tidal,2009PhRvD..80h4035D}:
\begin{equation}
    \Lambda (C, Y) = \frac{16}{15\Xi}(1 - 2C)^2[2 + 2C(Y - 1) - Y],
\end{equation}
where $\Xi$ is a function of $C$ and $Y$ given by
\begin{align}
\Xi (C, Y) &= 4C^3[13 - 11Y + C(3Y - 2) + 2C^2(1 + Y)] \nonumber \\
&\quad + 3(1 - 2C)^2[2 - Y + 2C(Y - 1)]\log(1 - 2C) \nonumber \\
&\quad + 2C[6 - 3Y + 3C(5Y - 8)].
\end{align}

The moment of inertia ($I$) of a NS is empirically related to $\Lambda$ via the function of \citep{yagi2017approximate} 
\begin{equation}
    \log_{10} \bar{I} = \sum_{n=0}^{4} a_n (\log_{10}\Lambda)^n,
    \label{eq:I-Lambda}
\end{equation}
where $a_{\rm n}=(0.65022,5.8594\times 10^{-2},5.1749\times 10^{-2},-3.6321\times 10^{-3},8.5909\times 10^{-5})$ are the fitting coefficients, and all variables are dimensionless ($\bar{I}=c^4I/G^2M^3$ and $G=c=1$). 

\begin{figure}[htb]
    \centering
    \gridline{\fig{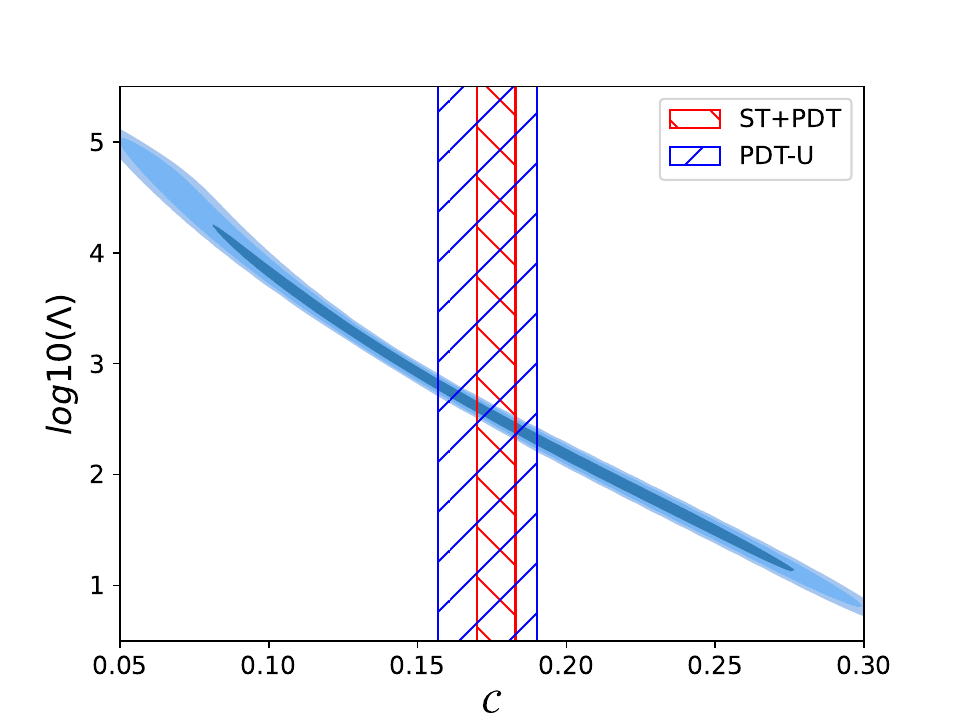}{0.32\textwidth}{(a)}
              \fig{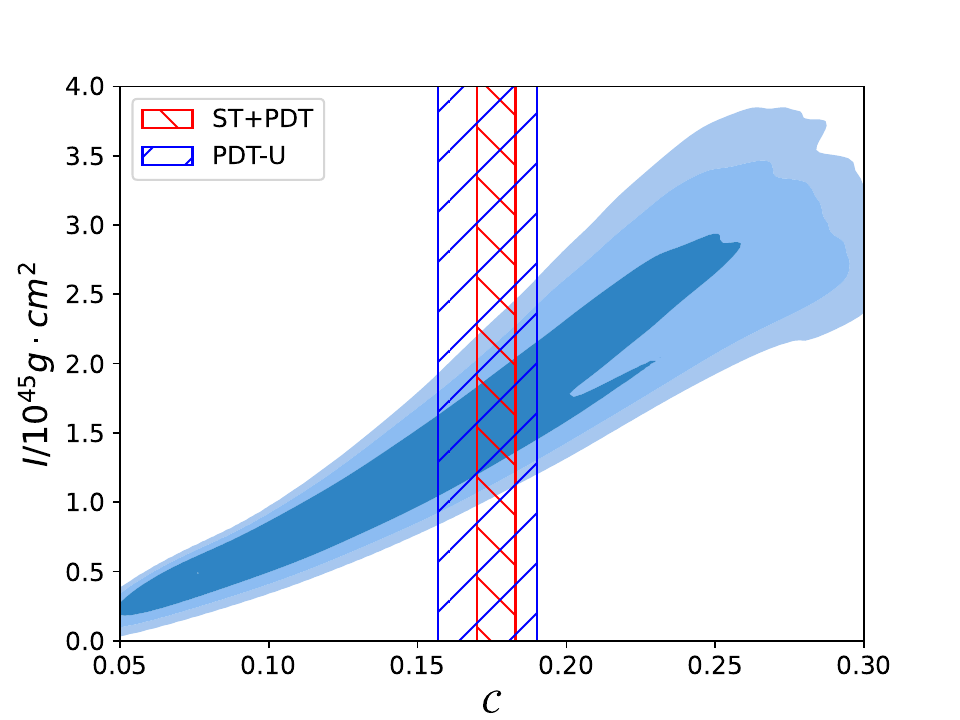}{0.32\textwidth}{(b)}
              \fig{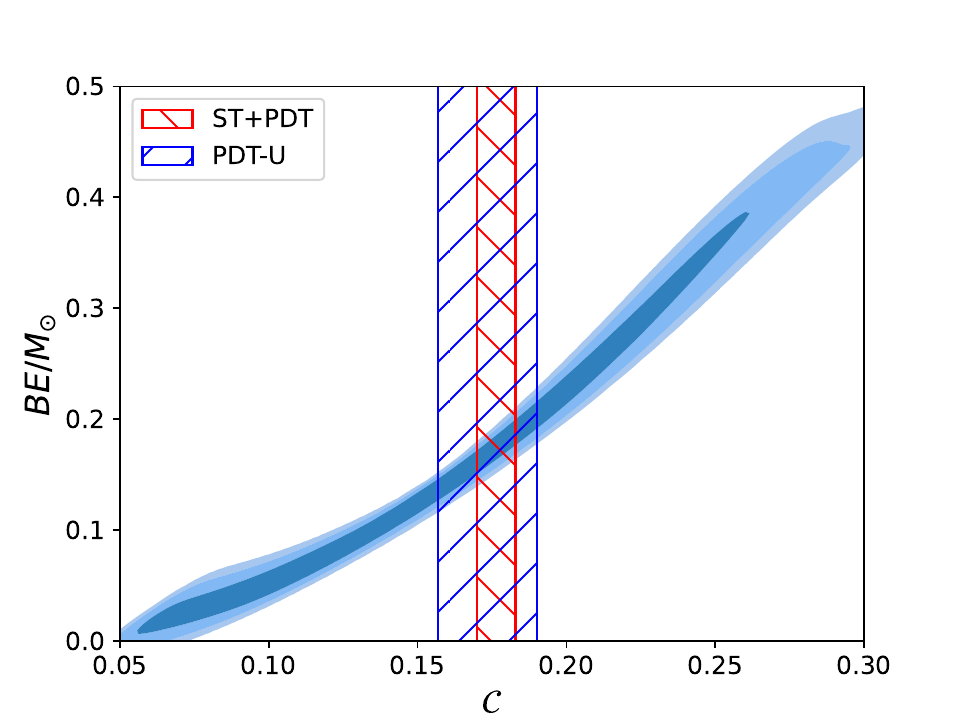}{0.32\textwidth}{(c)}}
    \caption{$(\Lambda, I, BE)$ as functions of compactness. The blue shaded regions, from light to dark, represent the result of GP model of $68\%$, $95\%$, and $99\%$ credible intervals, respectively. The red grid area indicates the $68\%$ credible interval of $C$ calculated by the ST+PDT model, and the blue grid area corresponds to the results of the PDT-U model.}
    \label{fig:C_relation}
\end{figure}

The gravitational binding energy (${\rm BE}$) can be calculated via \citep{steiner2016neutron}
\begin{equation}
    BE/M = \sum_{\rm n=0}^{4} b_{\rm n} \bar{I}^{-n},
\end{equation}
where $b_{\rm n}=(0.0075,~1.96_{-0.05}^{+0.05},~-12.8,~72,~-160_{-20}^{+20})$. 
In Fig.~\ref{fig:C_relation} we show the relationships of $\Lambda-{\cal C}$, $I-{\cal C}$ and ${\rm BE}-{\cal C}$, respectively. In principle, for a given ${\cal C}$ we can reasonably estimate the corresponding bulk properties of the NS. 

\section{Results}
\label{sec:results}
\begin{figure}[htb]
    \centering
    \includegraphics[width=0.6\textwidth]{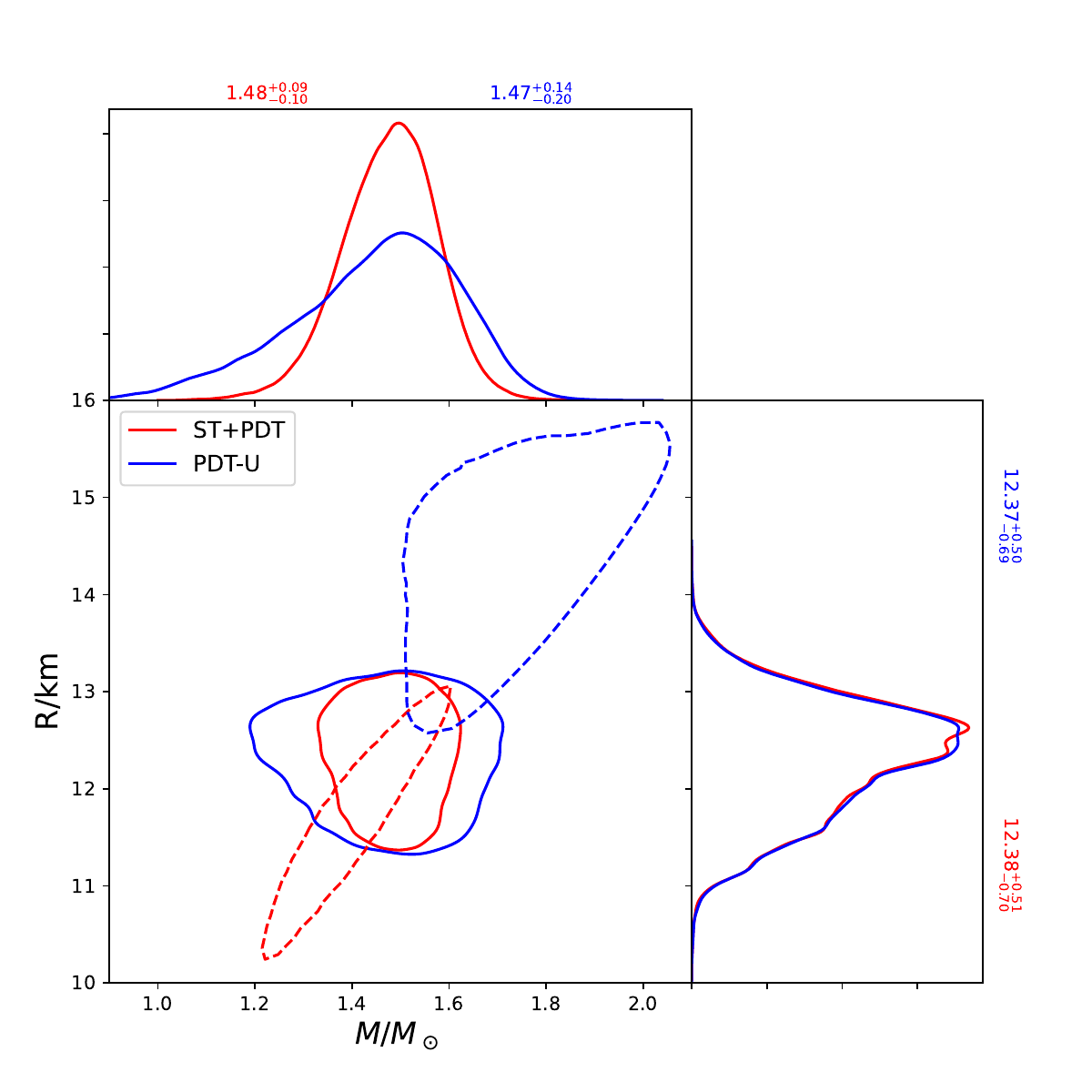}
    \caption{The solid lines are the estimation results for the mass and radius of PSR J0030+0451 with our approach (here we adopt the result with the GP approach). The dashed lines are the distribution of mass and radius directly inferred from the NICER$+$XMM-Newton X-ray data \citep{vinciguerra2024updated}. The lines in red are for the ST+PDT model and blue for the PDT-U model.}
    \label{fig:MR_comp}
\end{figure}

Incorporating the EoSs constrained by current data/information, we can reasonably infer the bulk properties of PSR J0030+0451.
Fig.~\ref{fig:MR_comp} shows our inferred mass/radius distributions. The calculation is carried out with EoS constraint result of the GP model (As shown in Fig.~\ref{fig:MC_compare}, the results of the three reconstruction methods are well consistent with each other). Different from the direct inference results only with the NICER and XMM-Newton data (i.e., the almost ``separated" counter regions marked by the dashed solid lines that are adopted from \citet{vinciguerra2024updated}), for the ST+PDT and PDT-U models our inferred mass and radius of PSR J0030+0451 are well consistent with each other. This is anticipated since both models yield rather similar ${\cal C}$ though the PDT-U model has larger error bar than that of the ST+PDT model, that is why the inferred mass for the former is more uncertain than the latter.
It is worth noting that since the radius is directly obtained from the constraints of the EoS, which is less influenced by the value of ${\cal C}$, our inferred radii in the ST+PDT and PDT-U models are rather similar.
Fig.~\ref{fig:MR_comp} also demonstrates that for PSR J0030+0451 our inferred $(M,~R)$ are closer to the NICER$+$XMM-Newton X-ray analysis under the ST+PDT model. We would also remark that our current results are also consistent with the initial mass/radius inference results with the X-ray data published in \citet{riley2019nicer} and \citet{miller2019psr}. Note that when adopting the same models used in \citet{riley2019nicer} and \citet{miller2019psr}, \citet{vinciguerra2024updated} found consistent results, although with more stringent inference requirements.  

\begin{figure}[htb]
    \centering
    \gridline{\fig{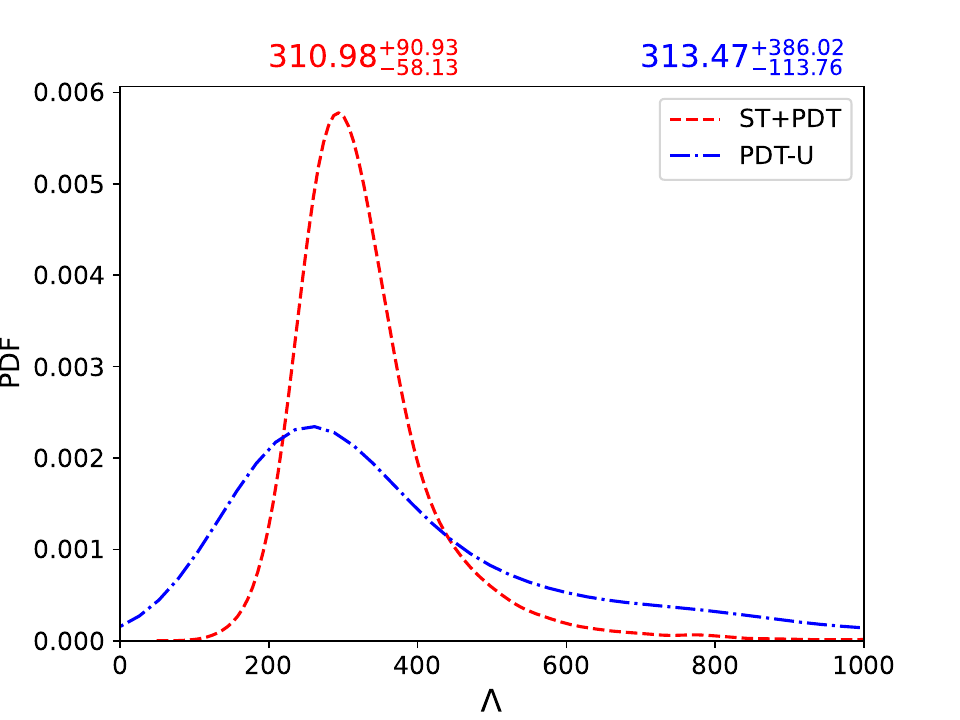}{0.32\textwidth}{(a)}
              \fig{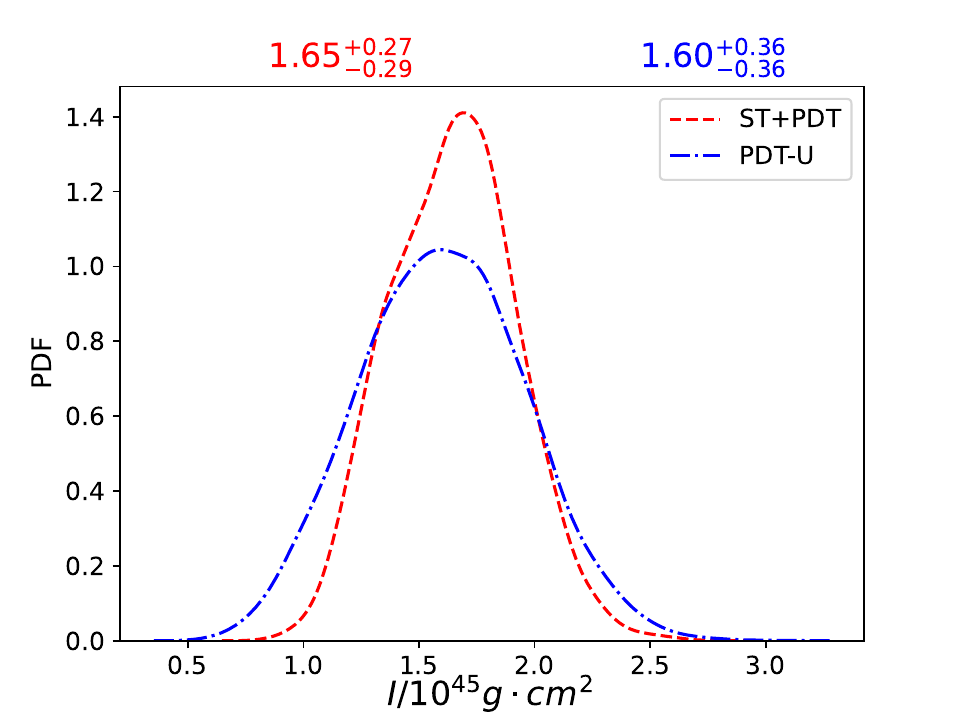}{0.32\textwidth}{(b)}
              \fig{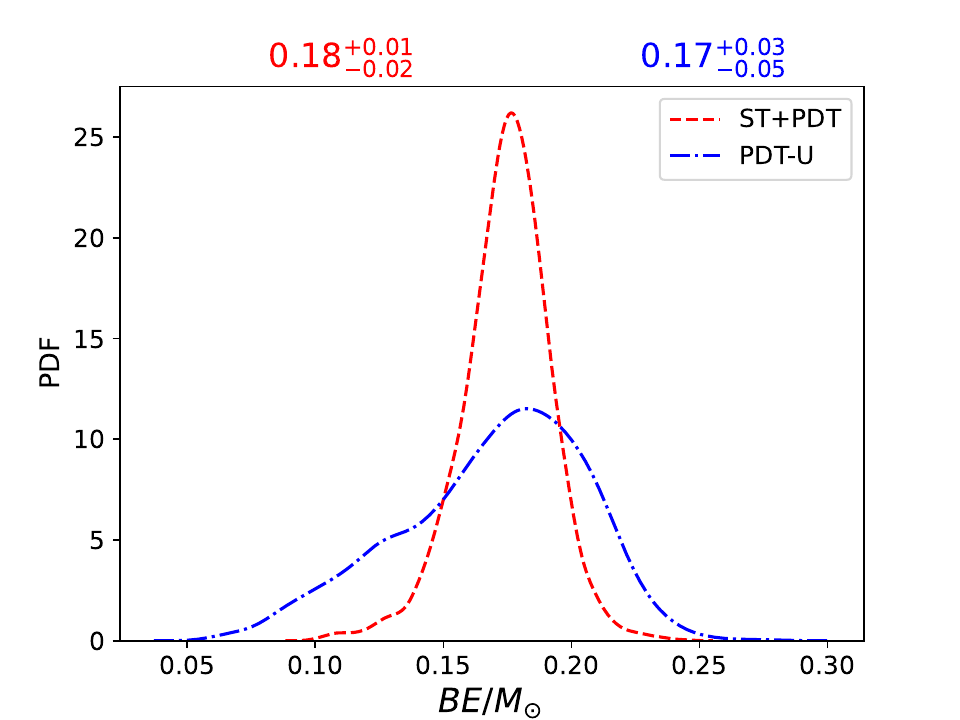}{0.32\textwidth}{(c)}}
    \caption{Estimations of tidal deformability $\Lambda$, the moment of inertia $I$, and gravitational binding energy ${\rm BE}$ of PSR J0030+0451. The colors of the two lines represent the two different models developed in the NICER$+$XMM-Newton X-ray data analysis, with ST+PDT model in red and PDT-U model in blue.}
    \label{fig:TIB_result}
\end{figure}

Now we estimate the tidal deformability, moment of inertia, and gravitational binding energy of PSR J0030+0451. The estimated  $\Lambda$ are $311^{+91}_{-58}$ and $313^{+386}_{-114}$ with the compactness found in the ST+PDT and PDT-U models (see Fig.~\ref{fig:TIB_result}(a)), which are consistent with each other. 
With the resulting tidal deformability, we can apply the empirical relationship (Equation \ref{eq:I-Lambda}) to obtain the distribution of the moment of inertia, with specific results shown in Fig.~\ref{fig:TIB_result}(b) (i.e., it is $1.65^{+0.27}_{-0.29} \times 10^{45}~{\rm g\cdot cm^{2}}$ and $1.60^{+0.36}_{-0.36}\times 10^{45}~{\rm g\cdot cm^{2}}$ for the ST+PDT and PDT-U model, respectively).
The estimated results for the gravitational binding energy ${\rm BE}$ are presented in Fig.~\ref{fig:TIB_result}(c), both results closely cluster around 0.18$M_{\odot}$. All of the above results are consistent with those presented by \citet{2019ApJ...885...39J} for an NS with a typical mass of $\sim1.4M_\odot$.

\section{Conclusion and Discussion}
\label{sec:discussion}
In recent years, there have been significant advancements in measuring the mass and radius of NSs. With the high signal-to-noise ratio gravitational wave signals of GW170817, people have simultaneously measured the mass and tidal deformability of a given neutron star for the first time. Since 2019, the NICER collaboration has published some measurement results of the mass and radius of the neutron stars with the X-ray data.
One of the target of NICER mission, PSR J0030+0451, is very close to Earth and has very bright X-ray radiation, and has long been considered an excellent target for measuring mass and radius using X-ray profile modeling method. However, the recent re-analysis \citep{vinciguerra2024updated} found that, based on more general assumptions and treatments, the mass and radius results for PSR J0030+0451 strongly depend on the adopted hot spot model. 
This means that the current high-precision X-ray data alone is still not sufficient to fully determine the mass and radius of the isolated NS PSR J0030+0451. Anyhow, the compactness (${\cal C}=M/R$) obtained from the two hot spot models (i.e., ST+PDT and PDT-U) developed in \citet{vinciguerra2024updated} are consistent with each other, despite that the error bars are different.

As shown in \citet{TangSP2020} and \citet{2022ApJ...930....4L}, for sources with known gravitational redshifts, it is possible to break the degeneracy between mass and radius based on the current multi-messenger data-constrained EoS. The physical basis is that the radius of NSs show very similar distribution over a wide range of masses, thus the gravitational redshift ($z_{\rm g}=1/\sqrt{1-2{\cal C}}-1$) is mainly determined by the NS mass. For sources like PSR J0030+0451 with known compactness ${\cal C}=M/R$, we can infer the mass and radius of a given compact object straightforwardly. In contrast to these two previous works, we do not solely use the parametric EoS reconstruction methods, and instead, we apply three (non-)parametric methods used in the literature, including the feed-forward neural network methods, piecewise linear sound speed methods, and Gaussian process methods. Now the NICER data for PSR J0030+0451 is excluded from the constraints on the NS EoS. To improve the precision of the constraints, we also incorporate the $M_{\rm TOV}$ information derived from the NS mass function by \citet{Fan2023} in our analysis. 
For the compactness obtained in the ST+PDT (PDT-U) model, we obtain $M=1.48^{+0.09}_{-0.10}M_\odot$ ($1.47^{+0.14}_{-0.20}M_\odot$), $R=12.38^{+0.51}_{-0.70}$ km ($12.37^{+0.50}_{-0.69}$ km). These results are closer to the NICER analysis results using the ST+PDT hot spot model (see Figure \ref{fig:MR_comp}). It is worth noting that the ${\cal C}-M$ relationship included in our analysis is almost independent of the EoS reconstruction algorithm (see Figure \ref{fig:MC_compare}), indicating that our results are model-insensitive. We also used the new constraints on the EoS to update the inferred mass-radius results for the three INSs with gravitational redshift measurements \citep{Hambaryan2014,Hambaryan2017}, finding little change (see Appendix \ref{app:ins-m} for the details) compared to the results of \citet{2022ApJ...930....4L}, which further suggests that our results are insensitive of the reconstruction method of the EoS. We further provided the $\Lambda-{\cal C}$, $I-{\cal C}$, and ${\rm BE}-{\cal C}$ relationship diagrams, allowing future sources with accurately measured ${\cal C}$ values to directly estimate their tidal deformability, moment of inertia, and gravitational binding energy. We also calculated the tidal deformability, moment of inertia, and gravitational binding energy for PSR J0030+0451. Finally, we would like to remark that PSR J0030+0451, due to its proximity, strong and stable X-ray emission, is one of the key targets for future facilities to measure the radius of NSs. Therefore, our mass-radius inference results are expected to be stringently tested in the future.

\acknowledgments
We thank Prof. Yi-Zhong Fan for stimulating discussion. This work is supported by the National Natural Science Foundation of China (No. 12233011,12073080, 11933010), the CAS Project for Young Scientists in Basic Research (No. YSBR-088), the General Fund of the China Postdoctoral Science Foundation (No. 2023M733735 and No. 2023M733736), and the Project for Special Research Assistant of the Chinese Academy of Sciences (CAS). J.L.J. acknowledges support by the Alexander von Humboldt Foundation and the ERC Advanced Grant ``JETSET: Launching, propagation and emission of relativistic jets from binary mergers and across mass scales'' (grant No. 884631).

\software{Bilby \citep[version 1.1.2;][\url{https://git.ligo.org/lscsoft/bilby/}]{2019ApJS..241...27A}, PyMultiNest \citep[version 2.6, ascl:1606.005, \url{https://github.com/JohannesBuchner/PyMultiNest}]{2016ascl.soft06005B}}

\bibliographystyle{aasjournal}
\bibliography{refer.bib}

\appendix
\section{The updated gravitational redshift$-$mass relationship}\label{app:zg-m}
\begin{figure}[htb]
    \centering
    \includegraphics[width=0.6\textwidth]{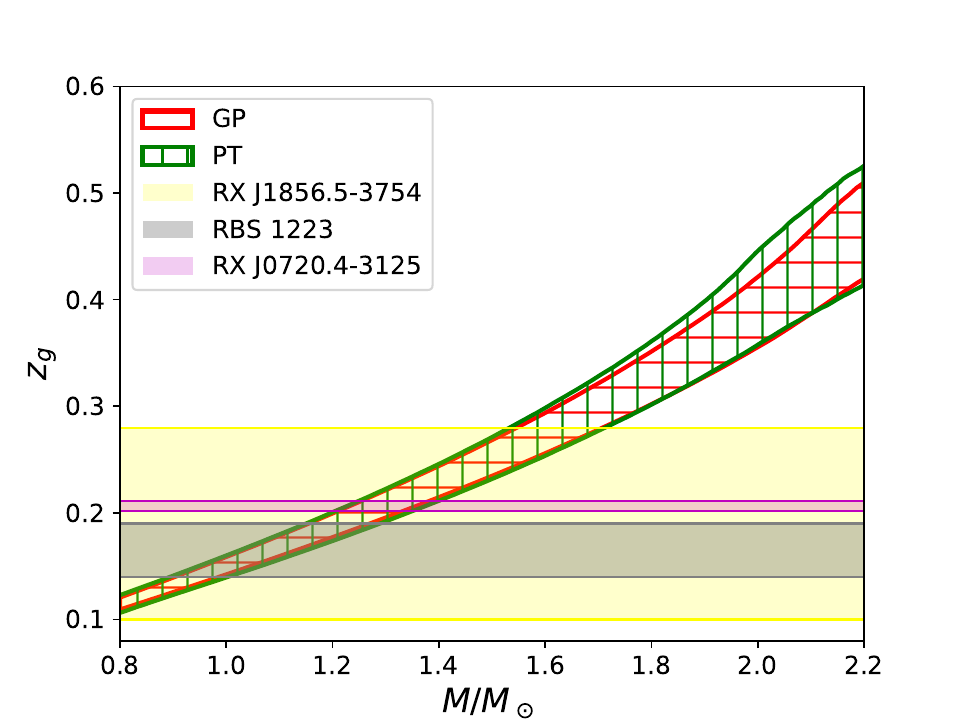}
    \caption{Relationship between the gravitational redshift and the mass of NS. Red- and green-band plots represent the $68\%$ credible regions of the results obtained from GP and phase transition \citep[PT;][]{2021PhRvD.104f3032T} models, respectively. The results of PT model is adopted from \citet{2022ApJ...930....4L}, derived utilizing the identical EoS inputs as those employed in the referenced study.} Grey-, purple-, and yellow-band plots show the $95\%$ highest-posterior-density interval of gravitational redshift measurements for sources RBS 1223, RX J0720.4-3125, and RX J1856.5-3754, respectively.
    \label{fig:zg_update}
\end{figure}

With the bounded EoSs found in Section \ref{sec:methods}, we now update the gravitational redshift-mass ($z_{\rm g}-M$) relationship. The results are shown in Fig.\ref{fig:zg_update}. 
It is rather similar to what we presented in Fig.~1 of \citet{2022ApJ...930....4L}, suggesting that our results are insensitive of the EoS reconstruction models (note that the EoS reconstruction approaches are very different from those adopted in \citet{2022ApJ...930....4L}) and the mass/radius data of PSR J0030+0451 alone does not significantly modify the EoS reconstruction results. The latter is also anticipated since the double NSs involved in GW170817 have similar masses/radii that can effectively constrain the EoS.

\section{Re-evaluation of the masses and radii of the isolated NS\MakeLowercase{s} with known $z_{\rm g}$}\label{app:ins-m}
\begin{figure}[htb]
    \centering
    \includegraphics[width=0.6\textwidth]{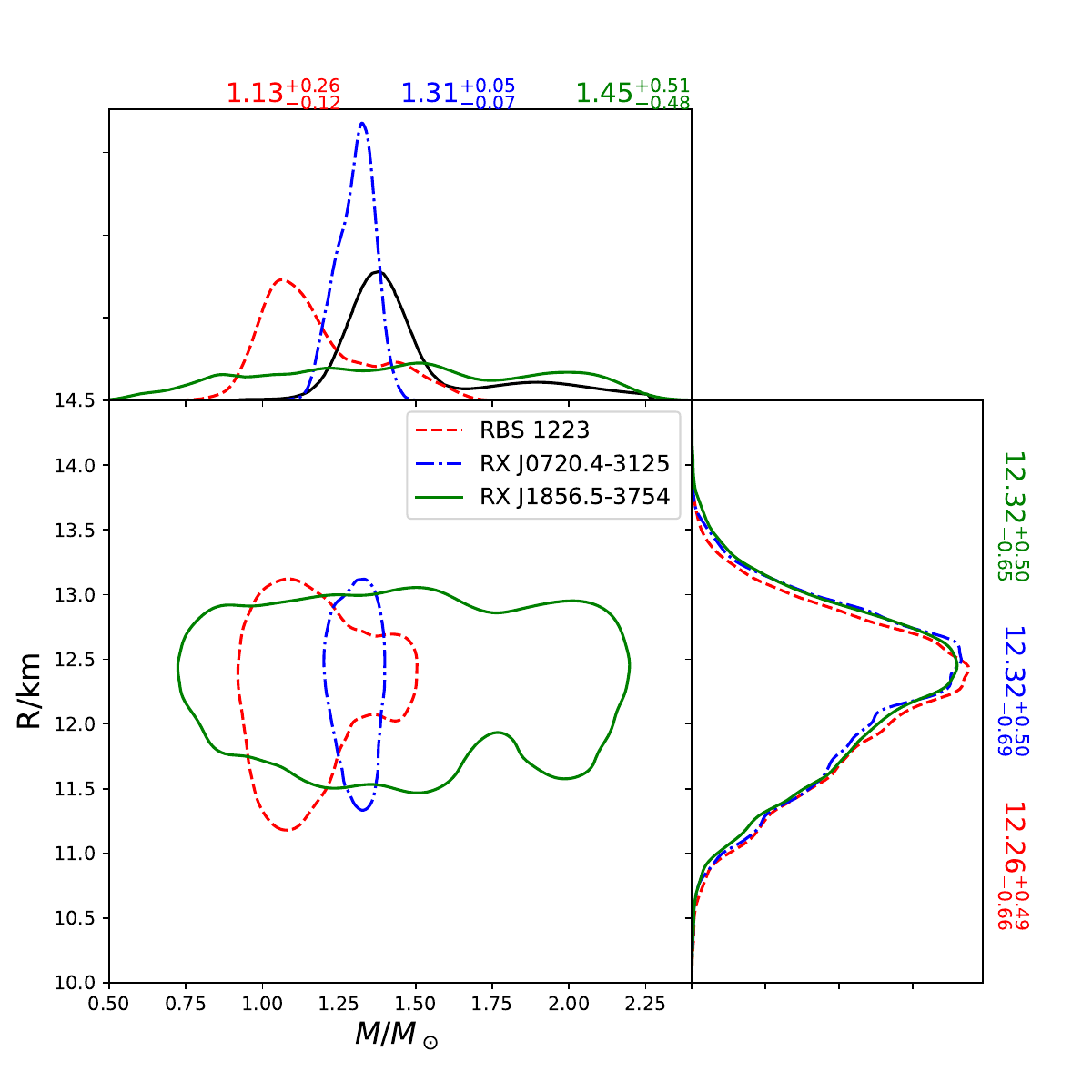}
    \caption{Re-estimation of the masses/radii of the three isolated NSs with measured gravitational redshifts subject to the newly constrained EoSs derived using the GP method. Previous estimates of these parameters are presented in Figure 2 of \citet{2022ApJ...930....4L}}. The solid black line represents the probability distribution of NS masses measured so far in binary systems \citep{Fan2023}.
    \label{fig:mr_update}
\end{figure}

We re-evaluate the masses and radii of the three INSs (including RBS 1223, RX J0720.4-3125, RX J1856.5-3754) with gravitational redshift measurements \citep{Hambaryan2014,Hambaryan2017}.  
Our results are reported in Figure \ref{fig:mr_update}. These mass results are very similar to those obtained in \citet{2022ApJ...930....4L}, indicating that the related results are insensitive to the reconstruction model of the equation of state and are credible. We observe that the mass-radius contour for RX J0720.4-3125 depicted in this study diverges from the results presented by \citet{2022ApJ...930....4L}, wherein their contour appears inclined. This discrepancy may be attributable to the fact that \citet{2022ApJ...930....4L} incorporated data from PSR J0030+0451, which was not considered in the current analysis. Furthermore, the measured gravitational redshift of the isolated NS RX J0720.4-3125 \citep[$z_g=0.205_{-0.003}^{+0.006}$;][]{Hambaryan2017} is remarkably similar to that of PSR J0030+0451 \citep[$z_g = 0.206_{-0.017}^{+0.014}$][]{riley2019nicer}. Consequently, RX J0720.4-3125 appears to reflect the mass-radius correlation reported for PSR J0030+0451 in the study by \citet{2022ApJ...930....4L}. The mass uncertainty obtained for RX J1856.5-3754 is large due to the high uncertainty in its gravitational redshift.

In the top part of Figure \ref{fig:mr_update}, we have also compared the mass distributions of the sources RBS 1223, RX J0720.4-3125, RX J1856.5-3754 with the probability distribution of  masses of the neutron stars measured in binary systems (i.e., the black solid line, which is adopted from \citet{Fan2023}). It is found that the masses of these INSs are relatively small (for RX J1856.5-3754, the error bar is quite large because of the poorly constrained $z_{\rm g}$), implying that the masses of (some) NSs in binary systems have been significantly increased through accretion.

\end{document}